\documentclass[12pt]{article}
\usepackage{graphicx}
\usepackage{epsfig}
\usepackage{epstopdf}
\DeclareGraphicsExtensions{.pdf,.eps,.png,.jpg,.mps}
\usepackage{caption}
\usepackage{float}
\usepackage{color}

\usepackage{url}
\usepackage[implicit=false]{hyperref}
\usepackage{cite}

\def\lsim{\mathrel{\rlap{\lower3pt\hbox{\hskip0pt$\sim$}}
     \raise1pt\hbox{$<$}}}         %less than or approx. symbol
\def\gsim{\mathrel{\rlap{\lower4pt\hbox{\hskip1pt$\sim$}}
     \raise1pt\hbox{$>$}}}         %greater than or approx. symbol

\usepackage{amsmath}
\usepackage{amsfonts}

\begin{document}
\begin{titlepage}

\centerline{\Large \bf ETF Risk Models}
\medskip

\centerline{Zura Kakushadze$^\S$$^\dag$\footnote{\, Zura Kakushadze, Ph.D., is the President of Quantigic$^\circledR$ Solutions LLC,
and a Full Professor at Free University of Tbilisi. Email: \href{mailto:zura@quantigic.com}{zura@quantigic.com}} and Willie Yu$^\sharp$\footnote{\, Willie Yu, Ph.D., is a Research Fellow at Duke-NUS Medical School. Email: \href{mailto:willie.yu@duke-nus.edu.sg}{willie.yu@duke-nus.edu.sg}}}
\bigskip

\centerline{\em $^\S$ Quantigic$^\circledR$ Solutions LLC}
\centerline{\em 680 E Main St, \#543, Stamford, CT 06901\,\,\footnote{\, DISCLAIMER: This address is used by the corresponding author for no
purpose other than to indicate his professional affiliation as is customary in
publications. In particular, the contents of this paper
are not intended as an investment, legal, tax or any other such advice,
and in no way represent views of Quantigic$^\circledR$ Solutions LLC,
the website \url{www.quantigic.com} or any of their other affiliates.
}}
\centerline{\em $^\dag$ Free University of Tbilisi, Business School \& School of Physics}
\centerline{\em 240, David Agmashenebeli Alley, Tbilisi, 0159, Georgia}
\centerline{\em $^\sharp$ Centre for Computational Biology, Duke-NUS Medical School}
\centerline{\em 8 College Road, Singapore 169857}
\medskip
\centerline{(September 5, 2021)}

\bigskip
\medskip

\begin{abstract}
{}We discuss how to build ETF risk models. Our approach anchors on i) first building a multilevel (non-)binary classification/taxonomy for ETFs, which is utilized in order to define the risk factors, and ii) then building the risk models based on these risk factors by utilizing the heterotic risk model construction of \cite{Het} (for binary classifications) or general risk model construction of \cite{HetPlus} (for non-binary classifications). We discuss how to build an ETF taxonomy using ETF constituent data. A multilevel ETF taxonomy can also be constructed by appropriately augmenting and expanding well-built and granular third-party single-level ETF groupings.
\end{abstract}
\medskip
\end{titlepage}

\newpage

\section{Introduction and Summary}

{}Exchange-traded funds (ETFs) can be thought of as an asset class in its own right.\footnote{\, For some literature on ETFs, see, e.g., \cite{Agapova2011a}, \cite{Aldridge2016}, \cite{Ben-David2017}, \cite{Bhattacharya2017}, \cite{Buetow2012}, \cite{Clifford2014}, \cite{Hill2015}, \cite{Krause2014}, \cite{Madhavan2012}, \cite{Madura2008}, \cite{Nyaradi2010}, \cite{Oztekin2017}.} There are around 2,500-2,600 ETFs listed in the U.S.\footnote{\, Strictly speaking, some of these are not ETFs but other exchange-traded products, such as exchange-trade notes (ETNs), etc. Depending on the context and end-user preferences, such non-ETF products can be included in or excluded from the universe we discuss below in the context of ETF risk model building. However, this will not be critical for our discussion below.\label{fn.ETF}} One of the allures of ETFs is their diversification power: ETFs allow an investor to gain exposure to different sectors, countries, asset classes, factors, etc., by taking positions in a relatively small number of instruments (i.e., ETFs) as opposed to taking positions in a large number of underlying instruments (e.g., thousands of stocks). On the other hand, the ETF universe itself has become rather diverse and in many cases a trading portfolio may contain a large number (in hundreds, if not larger) of ETFs. To manage risk in such a portfolio (e.g., via mean-variance optimization \cite{Markowitz1952}, by maximizing the Sharpe ratio \cite{Sharpe1994}, etc.), one needs to build a risk model for ETFs.

{}Just as in the case of stocks,\footnote{\, For a general discussion, see, e.g., \cite{GrinoldKahn}. For explicit implementations (including source code), see, e.g., \cite{HetPlus}, \cite{StatRM}.} modeling risk for $N$ ETFs involves forecasting the $N\times N$ covariance matrix for ETFs. Using the sample covariance matrix $C_{ij}$ ($i,j=1,\dots,N$) is not a viable option in most practical applications. This is because the number $N$ of ETF returns is much larger than the number $T$ of observations in the time-series of returns. The sample covariance matrix $C_{ij}$ in this case is badly singular: its rank is at best $T-1$. So, it cannot be inverted (which is required in, e.g., mean-variance optimization). Furthermore, the singularity of $C_{ij}$ is only a small part of the trouble: its off-diagonal elements (more precisely, sample correlations) are notoriously unstable out-of-sample (even if the matrix $C_{ij}$ is nonsingular).

{}So, just as in the case of stocks, we can forecast the ETF covariance matrix via a multifactor risk model, where ETF returns are (linearly) decomposed into contributions stemming from some number $K$ of common underlying factors plus idiosyncratic ``noise" pertaining to each ETF individually. This is a way of dimensionally reducing the problem in that one only needs to compute a factor covariance matrix $\Phi_{AB}$ ($A,B=1,\dots,K$), which is substantially smaller than $C_{ij}$ assuming $K\ll N$. The question then is, what should these common underlying factors be?

{}In Section \ref{sec2} we review how various types of risk factors are constructed for stocks. Then in Section \ref{sec3} we discuss what makes ETFs so different from stocks and why the multifactor risk model construction techniques for stocks cannot be readily adapted to ETFs. We also outline what can be done in the case of ETFs in this regard and discuss how to construct risk models for ETFs.\footnote{\, See Appendix \ref{app.B} for some important legalese.} We briefly conclude in Section \ref{sec4}.

\section{Multifactor Risk Models for Stocks}\label{sec2}

{}Given a time-series of stock returns $R_{is}$ (where $s=1,\dots, T$ labels the times (e.g., trading days) in the time-series) and the $N\times K$ factor loadings matrix $\Omega_{iA}$ (were $i=1,\dots, N$ labels stocks, while $A=1,\dots,K$ labels risk factors), one can construct a model covariance matrix $\Gamma_{ij}$ (which replaces the sample covariance matrix $C_{ij}$) using the explicit algorithm and source code given in \cite{HetPlus}. The factor loadings $\Omega_{iA}$ can be thought of (up to normalizations and/or factor rotations) as weights with which individual stocks contribute into the factors. Thus, the factor returns $f_{As}$ can be defined (again, up to normalizations and/or factor rotations) via\footnote{\, More precisely, given some generic factor loadings $\Omega_{iA}$, to satisfy the condition that the model variances $\Gamma_{ii}$ reproduce historical (or some other desired) variances, the actual factor loadings that enter the matrix $\Gamma_{ij}$ are given by $\Omega_{iA}/\gamma_i$, where the $N$-vectors $\gamma_i$ are fixed by the time-series $R_{is}$ and $\Omega_{iA}$ \cite{HetPlus}. For certain choices of $\Omega_{iA}$ things simplify and $\gamma_i \equiv 1$.}
\begin{equation}
 f_{As} = \sum_{i=1}^N \Omega_{iA}~R_{is}
\end{equation}
Defining the factor loadings matrix $\Omega_{iA}$ is then the key to building a risk model.

{}In the case of stocks, a priori reasonable choices for defining $\Omega_{iA}$ are as follows. In statistical risk models\footnote{\, See \cite{StatRM}, which gives complete source code, and references therein.} the factors are based on the first $K$ principal components of the sample covariance matrix $C_{ij}$ (or the sample correlation matrix).\footnote{\, The (often misconstrued) ``shrinkage" method \cite{Ledoit} is nothing but a special type of statistical risk models; see \cite{S=FM}, \cite{StatRM} for details.} In this case the number of factors is limited ($K < T-1$), and, furthermore, the principal components beyond the first one are inherently unstable out-of-sample. In contrast, factors based on a granular fundamental industry classification\footnote{\, E.g., BICS (Bloomberg Industry Classification System), GICS (Global Industry Classification Standard), ICB (Industry Classification Benchmark), SIC (Standard Industrial Classification), etc.} are much more ubiquitous (in hundreds), and also stable, as stocks seldom jump industries. Heterotic risk models \cite{Het}, \cite{HetPlus} based on such industry classifications sizably outperform statistical risk models. One can also include non-industry style factors, which are based on stocks' estimated/measured properties, e.g., size, value, growth, momentum, volatility, liquidity, etc. However, their number is limited, it is basically of order 10 or so.\footnote{\, The number of relevant style factors is even fewer in short-horizon risk models for use, e.g., in short-horizon quant trading applications \cite{4FM}, where it is 4 (or even fewer).} Furthermore, contrary to an apparent common misconception, style factors generally are poor proxies for modeling correlations and add little to no value \cite{HetPlus}.

{}This can be readily understood by noting that in factor models pair-wise correlations $\Psi_{ij}$ ($i\neq j$) are modeled via bilinear combinations
\begin{equation}
 \Psi_{ij} = \sum_{A,B = 1}^K \beta_{iA}~\Phi_{AB}~\beta_{jB}
\end{equation}
where $\beta_{iA} = \Omega_{iA}/\sigma_i$, and $\sigma_i^2 = \Gamma_{ii}$ are the variances. Here is a simple way to test whether a given style factor adds value. Let this style factor be $\beta_i$. Let $\nu_i\equiv 1$ be the unit vector. Then we can form 3 symmetrical bilinear combinations $x_{ij} = \nu_i\nu_j$, $y_{ij} = \nu_i\beta_j + \nu_j\beta_i$, and $z_{ij} = \beta_i\beta_j$. Let us further define a composite index $\alpha = \{(i,j)| i > j\}$. We can then define the following vectors (each with $N(N-1)/2$ components): $\Psi_\alpha$, $x_\alpha$, $y_\alpha$, $z_\alpha$. This is nothing but pulling a lower triangle of a matrix into a vector. We can now run a regression of $\Psi_\alpha$ over $x_\alpha$, $y_\alpha$, $z_\alpha$. Note that $x_\alpha$ is the intercept variable in this regression, and its coefficient is nothing but the average pair-wise correlation.\footnote{\, More precisely, this holds in the regression of $\Psi_\alpha$ over $x_\alpha$, demeaned $y_\alpha$, and demeaned $z_\alpha$.} For standard style factors, the other two variables $y_\alpha$ and $z_\alpha$ are generally poor explanatory variables for $\Psi_\alpha$, even in-sample \cite{HetPlus}. This is not surprising as a priori there is no reason whatsoever why, say, a bilinear tensor $\beta_i\beta_j$, where $\beta_i$ is, e.g., the log of the market cap, should be a good proxy for pair-wise correlations. The standard ``lore" for justifying using style factors as $\beta_i$ goes as follows. Suppose we take historical returns and regress them over some style factors. If the correlations are sufficiently high (e.g., if the regression coefficients have high annualized t-statistic \cite{FamaMacBeth}), then using such style factors in a multifactor model would appear to be justified. However, this argument has an evident caveat. A factor model -- by construction -- assumes that the residuals of that regression should have low correlations with the factor returns and the pair-wise correlations between different residuals should also be low. Furthermore, this would have to persist out-of-sample. None of this is guaranteed by simply having sufficiently high correlations between stock returns and style factor returns. In fact, in practice the above assumptions do not hold even in-sample, which is why style factors are poor proxies for pair-wise correlations.\footnote{\, Here the following clarifying remark is in order. Above we discussed regressing $\Psi_\alpha$ over $x_\alpha$, $y_\alpha$, $z_\alpha$, where $x_\alpha$ and $y_\alpha$ involve the unit vector $\nu_i$. This is not ad hoc. The factor loadings matrix should include the unit vector for a variety of reasons (including that it is subsumed in it once we add binary industry factors). In fact, for many style factors its presence is inevitable due to the ambiguity in their normalization. For instance, the log of the market cap style factor can be defined as $\beta_i = \ln(C_i/\mu)$, where $C_i$ is the market cap, and $\mu$ is some parameter with the dimensionality of market cap (i.e., it is measured in dollars). This $\mu$ is a priori arbitrary and its rescalings result in shifts of $\beta_i$ by a vector proportional to the unit vector $\nu_i$, so with $\beta_i$ we must also include $\nu_i$.}

{}So, based on the above discussion, it is reasonable to focus on risk factors based on a sufficiently granular and well-built fundamental industry classification. Another alternative is to replace the fundamental industry classification in the heterotic risk model construction by a statistical industry classification based on clustering (using machine learning techniques) the return time-series data \cite{StatIC},\footnote{\, Such statistical industry classifications can be constructed to be multilevel and granular.} without any reference to a fundamental industry classification. Risk models based on statistical industry classifications outperform statistical risk models but underperform risk models based on fundamental industry classifications \cite{StatIC}. Another approach based on machine learning is to aggregate a large number of heterotic risk models based on a single-level clustering (such as k-means \cite{Forgy}, \cite{Lloyd1957}, \cite{Lloyd1982}, \cite{Hartigan}, \cite{HartWong}, \cite{MacQueen}, \cite{Steinhaus}), which produces a non-factor risk model with the performance characteristics similar to those of risk models based on statistical industry classifications, so they still underperform risk models based on fundamental industry classifications \cite{MLRM}. Another issue with clustering the return time-series data is that tickers with stale prices (including illiquid stocks that do not trade much) can be easily misclassified by such algorithms with no alternative to fall back onto. In contrast, well-built fundamental industry classifications are based on analyzing companies' sources of revenues, products, services, customers, partners, suppliers, competitors, etc. (which information can be obtained for publicly traded companies from their 10-Q and 10-K SEC filings and other sources, even if their stocks are illiquid).

\section{ETFs Are Different from Equities}\label{sec3}

{}Considering that ETFs trade similarly to equities, it is tempting to treat them as such for the purposes of building ETF risk models. However, this would be suboptimal, to say the least. ETFs (with the caveat mentioned at fn. \ref{fn.ETF}), for the most part, are well-diversified portfolios of underlying instruments such as stocks, bonds, commodities, etc. The majority -- about 70\% -- of the ETFs listed in the U.S. are equity ETFs. Over 400 ETFs are bond ETFs, and about 100 ETFs are commodity ETFs. There is a sizable number of multi-asset ETFs, plus smaller numbers of ETFs such as real estate, alternatives, volatility, currency, etc. (depending on specific definitions). If all ETFs were equity based, things would have been much simpler. However, they are not. So, the question is, how to build a risk model for ETFs?

{}Since ETFs are baskets of their underlying instruments (a.k.a. constituents), it is tempting to build a risk model for ETFs based on a multi-asset risk model that covers all the constituents. There are two issues with this approach. First, one would have to either build such a multi-asset risk model or purchase an off-the-shelf offering (which is not cheap), not to mention the requisite constituent data, which can be quite painful to deal with. Furthermore, the number of constituents is much larger than the number of ETFs. Any multi-asset risk model has its own noise and imprecisions,\footnote{\, For instance, various issues with off-the-shelf equities risk models discussed in \cite{CusmtomRM}, the issues with style factors discussed above, etc., will all be inherited by the ETF risk model if the underlying equities risk model has them. Multi-asset risk models have even more noise and imprecisions than equities risk models precisely due to the mixing of different asset classes for which there is no clean analog of a fundamental industry classification for stocks. Etc.} which will all be propagated into the would-be ETF risk model. And the entire exercise would defy one of the beauties of ETFs, that they are already well-diversified portfolios that reduce the noise. It would be much simpler and nicer to build a risk model directly for ETFs, i.e., without using a huge multi-asset risk model for the constituents, similarly to how one would build one for stocks.

{}The problem is that there is no ``industry classification" for ETFs. Using principal components would produce a suboptimal model for the same reasons as those mentioned in Section \ref{sec2}. Using a clustering-based approach similar to the statistical industry classification for stocks is feasible for ETFs, but based on the discussion in Section \ref{sec2} also is expected to be suboptimal, with the added issue that many ETFs trade infrequently and classifying them using machine learning methods is challenging. Finally, a priori one could define several style factors for EFTs, e.g., those based on size, liquidity, volatility, momentum, earnings (for equities), dividends (for equities), duration (for bonds), credit rating (for bonds), etc. However, they would suffer from the same issues as those discussed in Section \ref{sec2}. Not only would they be poor proxies for pair-wise correlations, but they would also cover a mere fraction of the relevant risk space for the 2,500-2,600 ETFs the risk model should cover.

{}It is far more efficient to use a classification for ETFs, i.e., an ETF {\em taxonomy}. Such a classification can be binary, but need not be. In a binary classification we have a map $G$ between $N$ ETFs and their $K$ groupings (or clusters, categories, etc.)
\begin{equation}
 G:\{1,\dots,N\}\rightarrow\{1,\dots,K\}
\end{equation}
In fact, just as with stocks, this classification can be multilevel, so we can have another map $S$ that maps the $K$ groupings into a smaller number $F$ of their own groupings
\begin{equation}
 S:\{1,\dots,K\}\rightarrow\{1,\dots,F\}
\end{equation}
And there can be more than 2 such levels. In binary classifications each ETF belongs to one and only one grouping at each level, with multiple ETFs in each grouping.

{}Alternatively, we can have a non-binary classification where we still have $K$ groupings, but each ETF belongs to multiple such groupings with some weights $W_{iA}$. So, in the binary case for each value of $i$ we have only one nonzero $W_{iA}$, namely, $W_{iA} = \delta_{G(i), A}$. In the non-binary case for each value of $i$ we can have more than one nonzero $W_{iA}$, and we can normalize these weights to add up to 1, i.e.,
\begin{equation}
 \sum_{A=1}^K W_{iA} = 1
\end{equation}
Note that if the matrix $W_{iA}$ is not ``sparsely populated", then the $K$ factors will essentially become style-like factors. By ``sparsely populated" we mean that the majority of the elements of $W_{iA}$ are zero. More precisely, these can be small by some definition, i.e., if they are rounded within some reasonable tolerance, then the majority of $W_{iA}$ would be rounded to zero. As in the binary case, we could have a multilevel non-binary classification scheme. Note that for a binary classification we can use, e.g., a heterotic risk model construction \cite{Het}, \cite{HetPlus}. For a non-binary classification we can use the general risk model construction of \cite{HetPlus} with the factor loadings $\Omega_{iA}$ identified with $W_{iA}$ or the latter augmented with, e.g., the first principal components of blocks of the sample correlation matrix as in the heterotic risk model construction.\footnote{\, In the binary case the blocks are defined by the map $G$. In the non-binary case a priori there is no map $G$, only the weights $W_{iA}$. If these weights are ``sparsely populated", in some cases it might be possible to define an ``approximate" map $G$, which then defines the corresponding blocks.}

{}The challenge then is to build such multilevel (non-)binary classification schemes for ETFs. To appreciate this challenge, it is instructive to go through a {\em Gedankenexperiment} (thought experiment) of constructing the weights $W_{iA}$ based on an off-the-shelf multi-asset risk model. Such a risk model, as its risk factors, can have some industry factors (for stocks), some style factors (some of which may be specific to one asset class, while others may go across two or more asset classes), as well as some other binary or non-binary factors. Let us consider the industry factors. For a given equity ETF we would take the constituent weights (which are set to zero for the instruments in the multi-asset risk model that are not part of said ETF's basket) and compute the ETF's exposure to the industry risk factors. Generally, the resulting weights matrix $W_{iA}$ (with $A$ restricted to the industry factors) will not be ``sparsely populated". This is because many ETFs are not sector or industry specific. A similar situation would hold for, e.g., bond ETFs w.r.t. factors related to, e.g., duration, yield, credit rating, etc. The matrix $W_{iA}$ would be ``sparsely populated" only in the ``cross-asset" blocks, e.g., when $i$ corresponds to an equity ETF, while $A$ corresponds to a bond property such as duration. However, this does not make the entire matrix $W_{iA}$ ``sparsely populated", it only makes it -- and essentially trivially -- quasi-block-diagonal.\footnote{\, It is not block diagonal as there are multi-asset/other ETFs that have ``cross-asset" exposure.} The net result is that our weights matrix would effectively be populated with style-like factors, whose drawbacks we discussed in Section \ref{sec2}, including them being poor proxies for pair-wise correlations.

{}There are two choices. A classification can be built from scratch based on the constituent data. Alternatively, one can use an existing third-party grouping of ETFs and refine it, including by expanding it into a multilevel classification as such third-party groupings are single-level. The reason why a multilevel classification is required in the context of ETF risk model building is that the number of groupings in a single-level classification scheme that is granular enough is substantial (roughly, between 50 and 100), so the factor covariance matrix in a single-level heterotic risk model construction would be singular for short lookbacks (e.g., 21 trading days), and nonsingular but lacking out-of-sample stability for longer lookbacks (e.g., 252 trading days). Therefore, at least one more level with the number of groupings of order 10 is required for properly implementing the heterotic risk model construction.

{}Let us start with building a classification ground up from the constituent data. As mentioned above, equity ETFs are the largest subgroup of all ETFs. For equity ETFs we can start with a granular and well-built industry classification for stocks such as BICS or GIGS. We will focus on 3 levels: sectors (the least granular level), industries (a more granular level), and sub-industries (the most granular level). Let us start with sectors. Let $\omega_{ia}$ be the constituent weights, where as before $i=1,\dots,N$ labels ETFs, and $a=1,\dots,M$ labels stocks (recall that we are focused on equity ETFs for now). The weights are normalized such that
\begin{equation}
 \sum_{a=1}^M \omega_{ia} = 1
\end{equation}
Let $\Lambda_{aA} = 1$ if the stock labeled by $a$ belongs to the sector labeled by $A$ ($A=1,\dots,K_1$), and 0 otherwise.\footnote{\, This applies to a binary industry classification where each stock belongs to one and only one sector. In a non-binary classification, $\Lambda_{aA}$ can take fractional values interpreted as the weights with which the stock labeled by $a$ belongs to the sector labeled by $A$, such that $\sum_{A=1}^{K_1} \Lambda_{aA} = 1$.} The sector exposure can then be computed as
\begin{equation}\label{sector.exp}
 W_{iA} = \sum_{a=1}^M \omega_{ia}~\Lambda_{aA}
\end{equation}
There might be some stocks for which the industry classification is unavailable. We can simply omit such stocks from the sum in Eq. (\ref{sector.exp}) as such stocks are typically low cap, low liquidity issues. The exposures $W_{iA}$ computed this way will not be adequate for our purposes here. They need to be cleaned up. First, we can define a lower threshold $W_*$ (e.g., $W_* = 50\%$). If for a given ETF all exposures $W_{iA}$ are below $W_*$, then this ETF is not substantially concentrated in any sector and instead has broad exposure across sectors. Let us for now label such ETFs as ``broad" and we will deal with them below. For $W_* = 50\%$, a given ETF typically will have at most one $W_{iA} \geq W_*$.\footnote{\, In theory, the constituent weights $\omega_{ia}$ can ``conspire" such that there are two values of $W_{iA}$ exactly equal 50\%. In practice, however, such an occurrence is unlikely. To avoid this kind of ``ties", $W_*$ can be set to $50\% + \epsilon$, where $\epsilon$ is a small number.} We can then classify such an ETF into the sector labeled by $A$. If we decrease $W_*$, we can have more than one such sectors into which a given ETF can be classified and we could use the corresponding values of $W_{iA}$ to quantify the exposures to such sectors, while simply setting all $W_{iA} < W_*$ to 0. As $W_*$ is lowered, the matrix $W_{iA}$ becomes less and less ``sparsely populated" and starts to look more and more like a loadings matrix for style factors (which we wish to avoid).

{}Assuming a reasonable choice for $W_*$ and a reasonably ``sparsely populated" matrix $W_{iA}$ for all ETFs that have not been labeled as ``broad", a priori we could have too many ETFs classified into a given sector, i.e., if the number of such ETFs is greater than a predefined upper bound $N^*$ (e.g., $N^* = 30$). In this case, we would repeat the above procedure for classifying ETFs into sectors except that instead of sectors we would now classify them into industries. Such a classification may result in a meaningful splitting of sector ETFs into industry ETFs, but it may not. Thus, we might end up with mostly ``broad" industry ETFs and/or too few ETFs in each industry (or most industries), i.e., fewer than a predefined lower bound $N_*$ (e.g., $N_* = 3$). In this case splitting sectors into industries may not make sense. On the other hand, a priori we could end up with too large industries (with more than $N^*$ ETFs), which we could subsequently split into sub-industries using the same method as above, albeit in practice this usually is not expected to become a pressing issue.

{}So, using an industry classification for stocks, we can classify some equity ETFs into sectors, industries, etc., via the above procedure. Multi-asset ETFs that contain equities could also be classified this way in a non-binary fashion, i.e., their exposures $W_{iA}$ to, e.g., sectors can be computed this way assuming a reasonable choice of $W_*$. However, even for equity ETFs, we would end up having a large number of ``broad" ETFs, which we would need to classify further with added granularity. One way to do so is to use style-like factors to group ETFs into a reasonable number of groupings. E.g., for equity ETFs we can take the market cap $C_a$ for each constituent stock and compute the corresponding exposure for an ETF. There are more than one ways of doing this. Thus, we can define the corresponding exposures, call them $C_i$, via
\begin{equation}
 C_i = \sum_{a=1}^M \omega_{ia}~C_a
\end{equation}
However, considering that for stocks the size factor is defined using $\ln(C_a)$ (and not $C_a$), we can define the corresponding factor for the ETFs, call it ${\widetilde C}_i$, via
\begin{equation}
 {\widetilde C}_i = \sum_{a=1}^M \omega_{ia}~\ln(C_a)
\end{equation}
We can then define, e.g., 3 or more tranches for $C_i$ or ${\widetilde C}_i$ corresponding, e.g., to large cap, mid cap and small cap, or, alternatively and more granularly, to mega cap, large cap, mid cap, small cap and micro cap. An alternative way of constructing a cap based factor is to define the tranches for the market cap $C_a$ and then construct the exposures to these tranches as we did above for sectors. Thus, suppose we define $K_2$ tranches by defining $K_2 - 1$ values $C(1) < C(2) < \dots < C(K_2-1)$, where $C(1) > C(0)= 0$ and $C(K_2 - 1) < C(K_2) = \mbox{max}(C_a)$. So, the stock labeled by $a$ belongs to the tranche labeled by $A$ iff $C(A-1) < C_a \leq C(A)$, where $A=1,\dots, K_2$. This gives us a binary $M\times K_2$ matrix $\Theta_{aA}$. Using this binary matrix we can construct the exposures
\begin{equation}
 W_{iA} = \sum_{a=1}^M \omega_{ia}~\Theta_{aA}
\end{equation}
We can now apply the same method as for the sectors above by defining a lower threshold $W_*$. Depending on the value of $W_*$, we can get a binary or non-binary classification of ETFs according to the cap tranches with some number of ETFs classified as ``broad" or multi-cap. Similarly, we can classify ETFs into value, growth and ``blend" or ``mixed", depending on whether they are concentrated in value stocks, growth stocks, or a blend thereof. ETFs can be further classified by region (e.g., developed markets, emerging markets, frontier markets, and ``broad" or ``global"; developed markets can be further sub-classified, e.g., into North America, Europe and Asia-Pacific; emerging markets can be further subclassified, e.g., into Asia-Pacific, Europe, etc.), and further by country (or sub-region). Such classifications by region/country are applicable not only to equity ETFs but also to other ETFs, e.g., bond ETFs. As above for sector/industry classifications, upper $N^*$ and lower $N_*$ bounds on ETF counts within groupings can be applied to ensure that the groupings are not too large or fragmented. The exposures $W_{iA}$ can be binary or non-binary.

{}For bond ETFs, which are not as numerous as equity ETFs, there is no analog of the industry classification, market cap, or the growth/value factor. However, there are other factors, such as duration, credit rating, and bond type. Bond type can, e.g., be classified by Treasuries (U.S. Treasury bonds), TIPS (Treasury Inflation-Protected Securities), foreign sovereign bonds, municipal bonds (which can be further sub-classified, e.g., by state), mortgage-backed securities (MBS), corporate bonds, bank loans, convertible bonds, etc. Then we can compute the exposures $W_{iA}$, where $A$ labels such groupings, as we did above for sectors (by defining a lower threshold $W_*$, etc.), and ``broad" ETFs are essentially total bond market ETFs.

{}Credit rating can be classified as Investment-Grade bonds and High-Yield/Junk bonds. A given ETF can be classified according to these credit rating categories using the same method as above. Alternatively, one can take the actual (e.g., S\&P) credit rating $R_\alpha$ for the constituent bonds labeled by $\alpha$ ($\alpha=1,\dots,M_1$) and compute a given ETF's average credit rating $R_i$ via:
\begin{equation}\label{default.exp}
 R_i = \sum_{\alpha=1}^{M_1} \omega_{i\alpha}~R_\alpha
\end{equation}
where $\omega_{i\alpha}$ are the constituent weights, and numeric values for the bond credit rating $R_a$ must be assigned according to some scheme. A simple -- but not necessarily most accurate -- method is to use a linear scoring whereby AAA, AA, A, BBB, BB, B and below-B rated bonds are assigned numeric values $1,
2,3,4,5,6,7$ or, alternatively, $2,3,4,5,6,7,8$, respectively (see, e.g., \cite{Deng}). A more accurate method would be to use the actual default rates corresponding to each credit rating (see, e.g., \cite{Deng}; for default rates, see, e.g., \cite{Kraemer}), average the default rates via (\ref{default.exp}), and map $R_i$ back to the corresponding credit rating (by taking the closest credit rating in the pertinent credit-rating-to-default-rate-map). The default rate method is more accurate than the linear scoring method as the actual default rates increase nonlinearly with the aforesaid linear scores \cite{Deng}. At the end, for a given ETF we can get the corresponding average credit rating (which can take the same 21 values from AAA to C, if we use the S\&P credit rating system), which would be overly granular. These can now be grouped into less granular groupings, e.g., Investment-Grade (AAA through BBB-) and High-Yield (BB+ and below) groupings, or into a greater number of credit rating groupings (with more granularity).

{}For duration, the computation is more straightforward assuming the durations $T_\alpha$ of the constituents are known. We can compute a given ETF's average duration $T_i$ via
\begin{equation}
 T_i = \sum_{\alpha=1}^{M_1} \omega_{i\alpha}~T_\alpha
\end{equation}
Alternatively, as above, we can categorize the constituents by their duration, e.g., long (over 10 years), intermediate (3-10 years), short (1-3 years) and ultra-short (less than 1 year). This produces a binary matrix $\Delta_{\alpha A}$, where $A$ labels the above categories by duration, and the exposures are given by
\begin{equation}
 W_{iA} = \sum_{a=1}^{M_1} \omega_{i\alpha}~\Delta_{\alpha A}
\end{equation}
We can now apply the same method as for the sectors above by defining a lower threshold $W_*$. Depending on the value of $W_*$, we can get a binary or non-binary classification of ETFs according to the duration categories with some number of ETFs classified as ``broad" or all-duration/term. Some ETFs can be concentrated in bonds with a specific maturity year. These can be classified separately. However, further sub-classifying such ETFs by duration would result in an overly fragmented classification (i.e., small groupings). So we can classify such ETFs by duration.

{}As mentioned above, the next largest asset class by ETF count is commodities. They can be further classified by exposure to types of commodities, e.g., agricultural commodities, base metals, oil and gas, and precious metals (and a priori one can define other and/or more granular categories). We can now apply the same method as above by defining a lower threshold $W_*$. Depending on the value of $W_*$, we can get a binary or non-binary classification of commodity ETFs according to their exposure to the categories defined above with some number of ETFs classified as ``broad".

{}Next, let us discuss some of the smaller asset classes. There are about 50 real estate ETFs, which can be split into U.S. and global/international categories. There are about 20 currency ETFs and they can a priori be split further based on currency exposure, albeit the resulting categories would be small. Depending on a particular definition, the ``alternatives" (as an ``asset class") can contain ETFs based on hedge funds, long-short funds (e.g., dollar-neutral, 130/30, etc.), as well as other exposures (such as commodities, currencies, equities, multi-asset/diversified portfolios, etc.), and their treatment in constructing a multilevel ETF classification scheme requires some care (see below). Volatility ETFs also require care in the context of multilevel classifications (see below), including due to the fact that some of these can be leveraged and/or inverse ETFs. In fact, there are leveraged and/or inverse ETFs in other asset classes as well, including equities, bonds, commodities, currencies, real estate and multi-asset ETFs, so care is needed in their classification. Thus, leveraged ETFs can have higher volatility than their similar unleveraged counterparts, while inverse ETFs can be negatively correlated with theirs. In this regard, bunching, say, inverse ETFs together, without regard to their asset class, would make little sense as some asset classes can already be inversely correlated. This is yet another reason (along with the reason mentioned above) for building a multilevel classification.

{}Thus, we can have ETFs first classified according to their asset class at a less granular level and then subclassified according to other similarity criteria within their asset class. Classifying ETFs according to the asset class can be done by identifying the asset classes of the constituents and then applying the same method as above by defining a lower threshold $W_*$. This way, in most cases we will get a well-defined asset class for a given ETF, albeit there are also ``broad", i.e., multi-asset, ETFs, whose number can be around 100 depending on the definitions (including the value of $W_*$ -- see a detailed discussion of the method above). There are also some tricky cases, e.g., volatility ETFs, and ETFs with constituents that include ``hybrid" instruments such as preferred stocks and convertible bonds (see below).

{}Thus, in third-party single-level binary classifications, each ETF is classified into a single category, but some ETFs may be unclassified. E.g., this can happen with multi-asset ETFs, for which a more granular classification can be challenging. However, we cannot have N/A entries in the classification when constructing a heterotic risk model (see above). So, we must assign such ETFs to some categories. This can be done by comparing them with other multi-asset ETFs for which we do have non-N/A assignments. For instance, we can can use other attributes such as cap tranches (e.g., large cap, mid cap, small cap, etc. -- see above), style (value, growth and ``blend" -- see above), region, sub-region/country, etc. (see above). So, a given multi-asset ETF -- call it XYZ -- with the N/A category can be classified into the category with the largest number of multi-asset ETFs such that their aforesaid attributes (cap tranche, style, region, etc.) match those of XYZ. There can be ties in this process, which need to be resolved. I.e., a priori for XYZ we could get more than one category with matching attributes with the same number of multi-asset ETFs in each of these categories. We can resolve such ties by, e.g., picking the category that has most multi-asset ETFs (irrespective of the other attributes). In the unlikely event that there is still a tie, we can simply pick the final category at will (e.g., the first category in the list as it appears in the data) as such minutiae do not make a measurable difference considering all the other noise in the system.\footnote{\, There can be other asset classes for which ETFs with N/A categories can occur (e.g., volatility ETFs). In such rare cases a simple solution is to simply classify them as ``\{Asset Class\} -- Other", where ``\{Asset Class\}" is the place holder for the asset class (e.g., volatility).}

{}The next step is to build a multilevel classification starting with a granular single-level binary classification (which can be a third-party classification or an organically-built classification using the methodology described above). In fact, for our purposes here (i.e., for ensuring that the factor covariance matrix in the heterotic risk model construction is nonsingular -- see above) a 2-level classification will suffice. As the second, less granular level, we can take the asset class assignment. Let us assume that the first, granular level is binary (i.e., each ETF is classified into one and only one category). By definition, so is the second level as asset class assignments are binary (assuming the asset class categories include the multi-asset category -- see above). However, we also wish to have the property that each level-1 category corresponds to one and only one asset class. This may not be the case in, e.g., third-party classifications (while in organically-built classifications this can be ensured from the get-go). I.e., a priori we can have cases where two ETFs, call them ETF1 and ETF2, with the same category, call it C1, (erroneously or sloppily, as the case might be) assigned to them at the granular level, belong to different asset classes, call them AC1 and AC2. In these cases the easiest thing to do is to simply split the category C1 into two (or more) categories, call them C1.AC1, C1.AC2, etc., according to the asset classes AC1, AC2, etc., to which ETFs with category C1 belong. This way each resulting level-1 category C1.AC1, C1.AC2, etc., corresponds to one and only one asset class. One issue with this method is that it can produce many small level-1 categories, with only 1 or a few ETFs in each, which is overkill.

{}This can be mitigated as follows. Let $\Lambda_{iA} = \delta_{G(i), A}$ be an $N\times K$ matrix, where $G:\{1,\dots,N\}\mapsto \{1,\dots,K\}$ maps the $N$ ETFs (labeled by $i=1,\dots,N$) to the $K$ level-1 categories (labeled by $A=1,\dots,K$). Let $\Omega_{i\alpha} = \delta_{H(i), \alpha}$ be an $N\times F$ matrix, where $H:\{1,\dots,N\}\mapsto \{1,\dots,F\}$ maps the $N$ ETFs to the $F$ level-2 asset classes (labeled by $\alpha = 1,\dots,F$). Let $J(A) = \{i|G(i) = A\}$, i.e., $J(A)$ is the set of ETFs that belong to the level-1 category labeled by $A$. Let $Q(A) = \{\alpha|\Omega_{i\alpha} = 1, i\in J(A)\}$, i.e., for a given level-1 category labeled by $A$, $Q(A)$ is the set of asset classes (labeled by $\alpha$) of the ETFs in the level-1 category labeled by $A$. If, for a given value of $A$, we have $|Q(A)| = 1$ (where $|\cdot|$ denotes the number of elements of a set), then all is well (meaning, the level-1 category labeled by $A$ maps to one and only one asset class) and we do not need to do anything for the ETFs in this category. On the other hand, if, for a given value of $A$, we have $|Q(A)| > 1$, then, following the method described in the immediately preceding paragraph above, we would split the level-1 category labeled by $A$ according to the asset classes labeled by $\alpha\in Q(A)$. However, we wish to avoid ``over-splitting" (see above). Let $V_i$ be the average daily dollar volume (ADDV) for the ETF labeled by $i$ (where $V_i$ can be computed over the past 1 month, 3 months, 12 months, or some other suitable period).\footnote{\, If any $V_i$ is N/A, we can set it to zero.} Let
\begin{eqnarray}
 && V_{A\alpha} = \sum_{i \in J(A)} V_i~\Omega_{i\alpha}\\
 && V_A = \sum_{\alpha\in Q(A)} V_{A\alpha}\\
 && {\widetilde V}_{A\alpha} = V_{A\alpha} / V_A
\end{eqnarray}
Assuming $|Q(A)| > 1$, ${\widetilde V}_{A\alpha}$ measures the relative ADDV of different asset classes (labeled by $\alpha \in Q(A)$) within the same level-1 category. In many cases we will have one asset class with the lion's share of ADDV, with other asset classes representing small fractions. These are the cases we wish to avoid ``over-splitting". So, we can introduce a lower threshold, call it ${\widetilde V}_*$ (e.g., ${\widetilde V}_* = 0.1)$, and define ${\widetilde Q}(A) = \{\alpha | {\widetilde V}_{A\alpha} > {\widetilde V}_*, \alpha \in Q(A)\}$. If $|{\widetilde Q}(A)| = 1$, then instead of splitting the level-1 category labeled by $A$, we keep it intact and assign the asset class corresponding to $\alpha \in {\widetilde Q}(A)$ to all ETFs labeled by $i \in J(A)$. However, if $|{\widetilde Q}(A)| > 1$ we split the level-1 category labeled by $A$ following the method described in the immediately preceding paragraph above and deal with any resulting small categories using, e.g., the following method.

{}First, we define the smallest allowed size -- call it $n_*$ -- for level-1 categories (say, $n_* = 3$). Let us consider a given category labeled by $A$ with the number of ETFs $N_A < n_*$ (note that $N_A = |J(A)|$ -- see above). Let $P(A) = \{B|N_B \geq n_*, S(B) = S(A)\}$, where $S(\cdot)$ denotes the asset class (i.e., the level-2 category); i.e., $P(A)$ is the set of level-1 ETF categories (within the same asset class as the category labeled by $A$) with the number of ETFs at least as large as $n_*$. The idea is to reclassify the ETFs in the category labeled by $A$ into one or more categories in the set $P(A)$. One simple way to do this is to calculate pair-wise {\em serial} correlations $\rho_{iB} = \mbox{Cor}(R_{is}, R_{Bs})$ over some time period (e.g., 12 months, 3 months, or some other period) between daily returns -- as above, call them $R_{is}$ -- of the ETFs labeled by $i\in J(A)$ and the average category returns $R_{Bs}$, $B\in P(A)$, where we can define these average returns simply as follows:
\begin{equation}\label{cat-avg}
 R_{Bs} = \frac{1}{N^\prime(B, s)} \sum_{j\in J^\prime(B, s)} R_{js}
\end{equation}
where $N^\prime(B, s) = |J^\prime(B, s)|$, and, for a given value of $s$, $J^\prime(B, s)$ is the largest subset of $J(B)$ such that no $R_{js}$ is an N/A for $j\in J^\prime(B, s)$; i.e., for a given value of $s$ (which labels the trading day in the time-series), in the category labeled by $B$ we drop all ETFs with N/A returns and compute the average return $R_{Bs}$. We can then reassign the category for a given ETF labeled by $i\in A$ from $A$ to $B(i) = \mbox{argmax}_B~\rho_{iB}$; i.e., instead of classifying said ETF into the category labeled by $A$, we now classify it into the category $B$ for which the serial correlation of $R_{is}$ with $R_{Bs}$ is the largest.\footnote{\, In the context of heterotic risk models, for a given lookback $T$, we cannot have tickers with N/As in the time-series $R_{is}$, $s=1,\dots,T$. All tickers with N/As have to be dropped. N/As in $R_{is}$ can arise due to missing or bad data. E.g., we may wish to clean the data by not trusting large non-N/A values such that $|R_{is}| > R_*$, where $R_*$ is some predefined threshold (e.g., $R_* = 0.1$), and replace such values with N/As. To reduce the number of N/As, we can replace an N/A in $R_{is}$ by the corresponding level-1 category average $R_{As}$ (defined via (\ref{cat-avg})), where $A = G(i)$ (see above), i.e., $A$ labels the category to which the ETF labeled by $i$ belongs. In this regard, if we split level-1 categories based on some ADDV threshold ${\widetilde V}_*$ (see above), in theory we could end up with small categories with N/As for $R_{As}$ themselves. To avoid this, for the purposes of replacing N/As in $R_{is}$, we may wish to compute $R_{As}$ based on the original classification before such splitting is applied.\label{fn.ret.NA}}

{}Finally, let us mention that in some third-party classifications we may wish to split level-1 categories based on some more granular attributes. E.g., asset class ``Bond" ETFs may be grouped into level-1 categories that are not based on duration.\footnote{\, Another example would be credit rating.} Then we may wish to further split such categories based on duration, e.g., short, intermediate, long, all-duration (i.e., all-term), and also possibly unknown duration (if some bond ETFs have N/As for duration). We can then split bond ETFs using the same method (based on some ADDV threshold ${\widetilde V}_*$) as described above for splitting level-1 categories by asset class, except that here we split them by duration. Also, here we may wish to introduce a lower threshold, call it $m_*$, such that if the level-1 category has fewer than $m_*$ bond ETFs, then we do not spit it. E.g., $m_*$ can be set to the average number of ETFs per level-1 category (i.e., $m_* = N / K$, where $N$ is the number of ETFs, and $K$ is the number of level-1 categories before splitting them by duration). This way we can achieve greater granularity for bond ETFs. The same method can be applied to other asset classes and/or attributes, depending on a given third-party classification, its (level-1) categories and granularity, etc.

\section{Concluding Remarks}\label{sec4}

\begin{figure}[ht]
\centering
\includegraphics[scale=0.5]{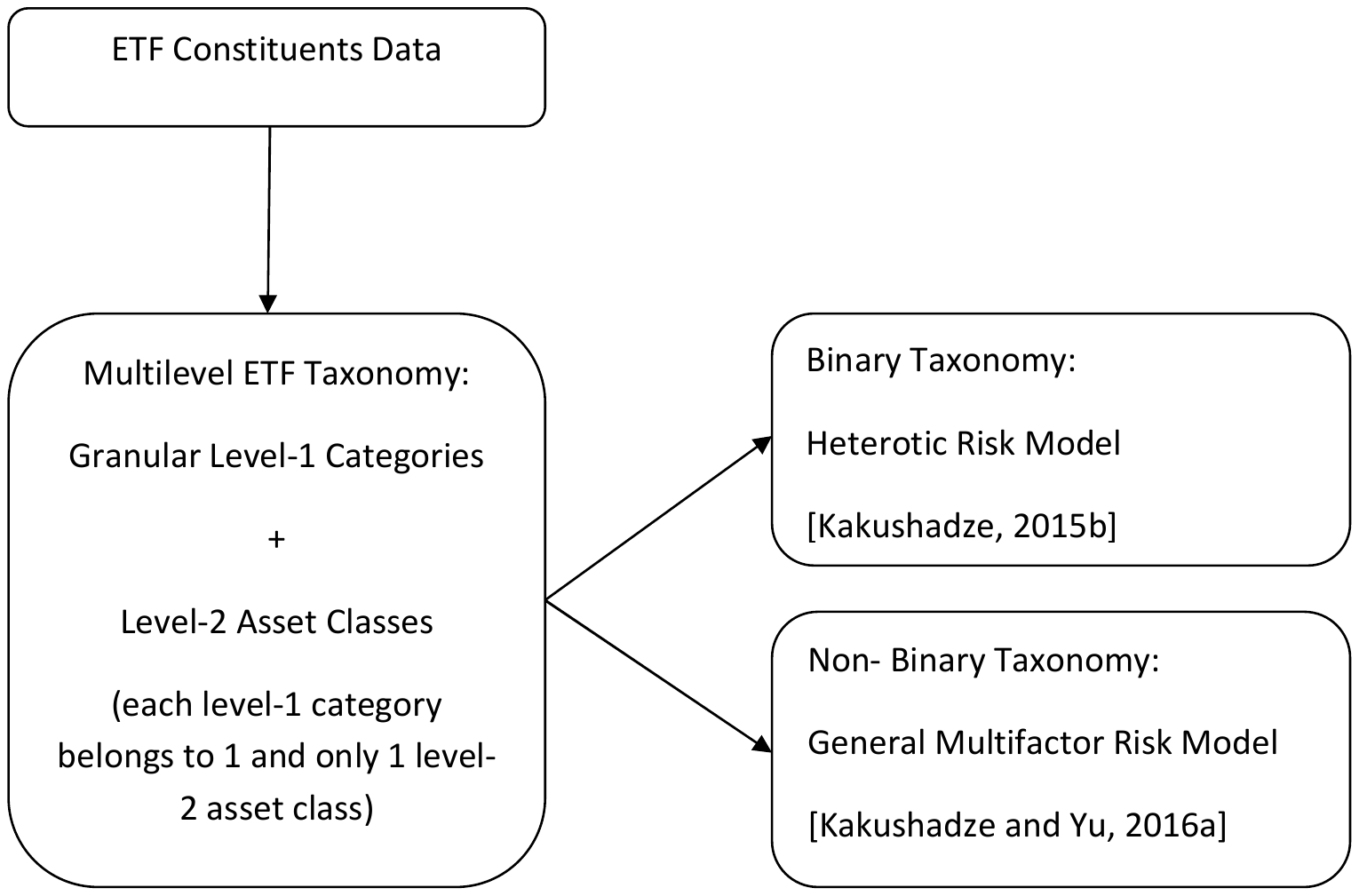}
\caption{A schematic depiction of the process of building ETF Risk Models starting from the ETF constituent data, building a multilevel (e.g., 2-level) ETF taxonomy, which can be binary or non-binary, and then applying the Heterotic Risk Model construction \cite{Het} (in the binary taxonomy case) or the general multifactor risk model construction \cite{HetPlus} (in the non-binary taxonomy case).}
\label{Fig1}
\end{figure}

\begin{figure}[ht]
\centering
\includegraphics[scale=0.5]{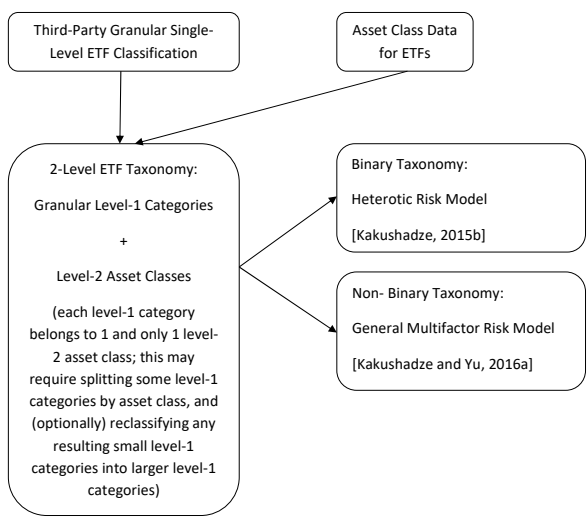}
\caption{A schematic depiction of the process of building ETF Risk Models starting from a third-party granular single-level ETF classification, augmenting it with ETF asset class data to build a 2-level ETF taxonomy, which can be binary or non-binary, and then applying the Heterotic Risk Model construction \cite{Het} (in the binary taxonomy case) or the general multifactor risk model construction \cite{HetPlus} (in the non-binary taxonomy case). }
\label{Fig2}
\end{figure}

{}Let us briefly conclude with some remarks to help solidify our discussion above. The organic ETF risk model construction described above, which combines a multilevel ETF classification with the heterotic risk model construction (see above) is very different from off-the-shelf risk models that cover ETFs. The latter, apart from limited coverage (which, to be fair, could partly be ``justified" based on liquidity considerations  -- see below), unlike our ETF risk models, are not built based on a binary (or non-binary but sparsely-populated) classification matrix $W_{iA}$, using which, as we argue above, makes more sense than style (or style-like non-sparsely-populated) factors. Also, while a binary classification could be built using machine learning (clustering) methods (see above), we are not doing this here as such classifications tend to be relatively unstable out-of-sample (see above).\footnote{\, One may argue that we do use subtle ``machine learning" techniques, e.g., when we reclassify ETFs from overly small categories to larger categories based on correlations (see above); however, apart from possibly being just ``buzz-wordy" semantics, this is a far cry from clustering-based machine learning classification algorithms (see \cite{StatIC}) we discuss above.} Instead, we define risk factors based on fundamental data as input. These factors can be built organically based on the constituent data using the methods described above. Alternatively, they can be defined using a well-built and granular third-party classification for defining level-1 categories and augmenting them with asset classes as level-2 categories (with some level-1 categories properly split to have one and only one asset class for each level-1 category -- see above).\footnote{\, In Quantigic$^\circledR$ ETF Risk Models we use a proprietary 2-level classification utilizing a granular third-party classification as level-1 categories further spit by asset class, with the resulting small categories reclassified as discussed above. See Fig.\ref{Fig1} and Fig.\ref{Fig2} for schematic processes when a 2-level classification is built organically and using a third-party granular classification, respectively.} Finally, let us mention that many ETFs are illiquid and do not trade much. The beauty of our approach is that we can still cover such ETFs so long as they are covered by the ETF classification. Then N/As in their return time-series can be replaced, e.g., as in fn. \ref{fn.ret.NA}, so the heterotic risk model can be constructed. Alternatively, one may wish to take the view (which may or may not be acceptable/desirable depending on a particular application of the ETF risk model) that such ETFs are too illiquid and modeling risk for them is ``meaningless". Either way, it is good to have an option to include such ETFs in a risk model, as they can always be excluded by restricting the ETF universe to a subset, keeping in mind that for illiquid ETFs category-average returns are used.

\appendix

\section{DISCLAIMERS}\label{app.B}

{}Wherever the context so requires, the masculine gender includes the feminine and/or neuter, and the singular form includes the plural and {\em vice versa}. The author of this paper (``Author") and his affiliates including without limitation Quantigic$^\circledR$ Solutions LLC (``Author's Affiliates" or ``his Affiliates") make no implied or express warranties or any other representations whatsoever, including without limitation implied warranties of merchantability and fitness for a particular purpose, in connection with or with regard to the content of this paper including without limitation any code or algorithms contained herein (``Content").

{}The reader may use the Content solely at his/her/its own risk and the reader shall have no claims whatsoever against the Author or his Affiliates and the Author and his Affiliates shall have no liability whatsoever to the reader or any third party whatsoever for any loss, expense, opportunity cost, damages or any other adverse effects whatsoever relating to or arising from the use of the Content by the reader including without any limitation whatsoever: any direct, indirect, incidental, special, consequential or any other damages incurred by the reader, however caused and under any theory of liability; any loss of profit (whether incurred directly or indirectly), any loss of goodwill or reputation, any loss of data suffered, cost of procurement of substitute goods or services, or any other tangible or intangible loss; any reliance placed by the reader on the completeness, accuracy or existence of the Content or any other effect of using the Content; and any and all other adversities or negative effects the reader might encounter in using the Content irrespective of whether the Author or his Affiliates is or are or should have been aware of such adversities or negative effects.

\newpage\clearpage

\newpage\clearpage

\end{document}